\documentclass[twocolumn,superscriptaddress,amsmath,amssymb,showpacs]{revtex4-1}

\usepackage{graphicx}
\usepackage{epsfig}
\usepackage{hyperref}
\usepackage{dcolumn}
\usepackage{amssymb}
\usepackage{amsmath}
\usepackage{natbib}

\newcommand{\PolV}{
 {\mathchoice
  {\vcenter{\hbox{\includegraphics[height=2ex,width=2ex]{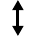}}}}
  {\vcenter{\hbox{\includegraphics[height=2ex,width=2ex]{flecha.pdf}}}}
  {\vcenter{\hbox{\includegraphics[height=1.6ex,width=1.6ex]{flecha.pdf}}}}
  {\vcenter{\hbox{\includegraphics[height=1.2ex,width=1.2ex]{flecha.pdf}}}}
 }
}

\newcommand{\PolH}{
 {\mathchoice
  {\vcenter{\hbox{\includegraphics[angle=90,width=2ex]{flecha.pdf}}}}
  {\vcenter{\hbox{\includegraphics[angle=90,width=2ex]{flecha.pdf}}}}
  {\vcenter{\hbox{\includegraphics[angle=90,width=1.6ex]{flecha.pdf}}}}
  {\vcenter{\hbox{\includegraphics[angle=90,width=1.2ex]{flecha.pdf}}}}
 }
}

\newcommand{\PolD}{
 {\mathchoice
  {\vcenter{\hbox{\includegraphics[angle=-45,width=2ex]{flecha.pdf}}}}
  {\vcenter{\hbox{\includegraphics[angle=-45,width=2ex]{flecha.pdf}}}}
  {\vcenter{\hbox{\includegraphics[angle=-45,width=1.6ex]{flecha.pdf}}}}
  {\vcenter{\hbox{\includegraphics[angle=-45,width=1.2ex]{flecha.pdf}}}}
 }
}
\newcommand{\Pold}{
 {\mathchoice
  {\vcenter{\hbox{\includegraphics[angle=45,width=2ex]{flecha.pdf}}}}
  {\vcenter{\hbox{\includegraphics[angle=45,width=2ex]{flecha.pdf}}}}
  {\vcenter{\hbox{\includegraphics[angle=45,width=1.6ex]{flecha.pdf}}}}
  {\vcenter{\hbox{\includegraphics[angle=45,width=1.2ex]{flecha.pdf}}}}
 }
}

\newcommand{\PolL}{
 {\mathchoice
  {\vcenter{\hbox{\includegraphics[height=1.9ex]{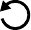}}}}
  {\vcenter{\hbox{\includegraphics[height=1.9ex]{flechaL.pdf}}}}
  {\vcenter{\hbox{\includegraphics[height=1.5ex]{flechaL.pdf}}}}
  {\vcenter{\hbox{\includegraphics[height=1.1ex]{flechaL.pdf}}}}
 }
}

\newcommand{\PolR}{
 {\mathchoice
  {\vcenter{\hbox{\includegraphics[height=1.9ex]{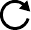}}}}
  {\vcenter{\hbox{\includegraphics[height=1.9ex]{flechaR.pdf}}}}
  {\vcenter{\hbox{\includegraphics[height=1.5ex]{flechaR.pdf}}}}
  {\vcenter{\hbox{\includegraphics[height=1.1ex]{flechaR.pdf}}}}
 }
}

\newcommand{\Ket}[1]{
\left | #1 \right >
}

\newcommand{\TEMV}{
 {\mathchoice
  {\vcenter{\hbox{\includegraphics[height=2ex]{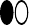}}}}
  {\vcenter{\hbox{\includegraphics[height=2ex]{TEM.pdf}}}}
  {\vcenter{\hbox{\includegraphics[height=1.6ex]{TEM.pdf}}}}
  {\vcenter{\hbox{\includegraphics[height=1.2ex]{TEM.pdf}}}}
 }
}

\newcommand{\TEMH}{
 {\mathchoice
  {\vcenter{\hbox{\includegraphics[angle=90,width=2ex]{TEM.pdf}}}}
  {\vcenter{\hbox{\includegraphics[angle=90,width=2ex]{TEM.pdf}}}}
  {\vcenter{\hbox{\includegraphics[angle=90,width=1.6ex]{TEM.pdf}}}}
  {\vcenter{\hbox{\includegraphics[angle=90,width=1.2ex]{TEM.pdf}}}}
 }
}

\newcommand{\TEMD}{
 {\mathchoice
  {\vcenter{\hbox{\includegraphics[angle=45,width=2.4ex]{TEM.pdf}}}}
  {\vcenter{\hbox{\includegraphics[angle=45,width=2.4ex]{TEM.pdf}}}}
  {\vcenter{\hbox{\includegraphics[angle=45,width=1.9ex]{TEM.pdf}}}}
  {\vcenter{\hbox{\includegraphics[angle=45,width=1.45ex]{TEM.pdf}}}}
 }
}
\newcommand{\TEMd}{
 {\mathchoice
  {\vcenter{\hbox{\includegraphics[angle=135,width=2.4ex]{TEM.pdf}}}}
  {\vcenter{\hbox{\includegraphics[angle=135,width=2.4ex]{TEM.pdf}}}}
  {\vcenter{\hbox{\includegraphics[angle=135,width=1.9ex]{TEM.pdf}}}}
  {\vcenter{\hbox{\includegraphics[angle=135,width=1.45ex]{TEM.pdf}}}}
 }
}

\newcommand{\TEML}{
 {\mathchoice
  {\vcenter{\hbox{\includegraphics[height=2ex]{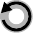}}}}
  {\vcenter{\hbox{\includegraphics[height=2ex]{TEML.pdf}}}}
  {\vcenter{\hbox{\includegraphics[height=1.6ex]{TEML.pdf}}}}
  {\vcenter{\hbox{\includegraphics[height=1.2ex]{TEML.pdf}}}}
 }
}

\newcommand{\TEMR}{
 {\mathchoice
  {\vcenter{\hbox{\includegraphics[height=2ex]{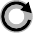}}}}
  {\vcenter{\hbox{\includegraphics[height=2ex]{TEMR.pdf}}}}
  {\vcenter{\hbox{\includegraphics[height=1.6ex]{TEMR.pdf}}}}
  {\vcenter{\hbox{\includegraphics[height=1.2ex]{TEMR.pdf}}}}
 }
}

\newcommand{\mc}[3]{\multicolumn{#1}{#2}{#3}}
\AtBeginDocument{%
  
}
\usepackage[pdftex]{color}
\usepackage{footnote}  % Footnotes on tables, using savenotes
\usepackage{multirow} % Para hacer col spanning
\usepackage{xfrac}
\usepackage{placeins} % Para usar \FloatBarrier

\newcommand{\etal}{\textit{et al.}\,}

\usepackage{amsmath}
\usepackage{bbm,amsfonts}

\usepackage{amsthm}
\theoremstyle{definition}

\usepackage{xy}
\xyoption{matrix}
\xyoption{frame}
\xyoption{arrow}
\xyoption{arc}

\usepackage{ifpdf}
\ifpdf
\else
\PackageWarningNoLine{Qcircuit}{Qcircuit is loading in Postscript mode.  The Xy-pic options ps and dvips will be loaded.  If you wish to use other Postscript drivers for Xy-pic, you must modify the code in Qcircuit.tex}
\xyoption{ps}
\xyoption{dvips}
\fi

\entrymodifiers={!C\entrybox}

\newcommand{\ket}[1]{{\left\vert{#1}\right\rangle}}
\newcommand{\qw}[1][-1]{\ar @{-} [0,#1]}
\newcommand{\qwx}[1][-1]{\ar @{-} [#1,0]}
\newcommand{\gate}[1]{*+<.6em>{#1} \POS ="i","i"+UR;"i"+UL **\dir{-};"i"+DL **\dir{-};"i"+DR **\dir{-};"i"+UR **\dir{-},"i" \qw}
\newcommand{\control}{*!<0em,.025em>-=-<.2em>{\bullet}}
\newcommand{\controlo}{*+<.01em>{\xy -<.095em>*\xycircle<.19em>{} \endxy}}
\newcommand{\ctrl}[1]{\control \qwx[#1] \qw}
\newcommand{\ctrlo}[1]{\controlo \qwx[#1] \qw}
\newcommand{\multigate}[2]{*+<1em,.9em>{\hphantom{#2}} \POS [0,0]="i",[0,0].[#1,0]="e",!C *{#2},"e"+UR;"e"+UL **\dir{-};"e"+DL **\dir{-};"e"+DR **\dir{-};"e"+UR **\dir{-},"i" \qw}
\newcommand{\ghost}[1]{*+<1em,.9em>{\hphantom{#1}} \qw}
\newcommand{\rstick}[1]{*!L!<-.5em,0em>=<0em>{#1}}
\newcommand{\lstick}[1]{*!R!<.5em,0em>=<0em>{#1}}
\newcommand{\Qcircuit}{\xymatrix @*=<0em>}
\newcommand{\beqa}{\begin{eqnarray}}
\newcommand{\eeqa}{\end{eqnarray}}
\newcommand{\beq}{\begin{equation}}
\newcommand{\eeq}{\end{equation}}



\begin{document}
\title{Manipulating Transverse Modes of Photons for Quantum Cryptography}
% \pacs{03.65.Wj,03.67.Mn,42.50.Dv,42.65.Lm}

\author{Marcelo Alejandro \surname{Luda}}
  \affiliation{Departamento de F\'{\i}sica, FCEyN, UBA, Pabell\'on 1, Ciudad Universitaria, 1428 Buenos Aires, Argentina}
  \affiliation{CEILAP, CITEDEF, J.B. de La Salle 4397, 1603 Villa Martelli, Buenos Aires, Argentina}

\author{Miguel Antonio \surname{Larotonda}}
  \affiliation{Departamento de F\'{\i}sica, FCEyN, UBA, Pabell\'on 1, Ciudad Universitaria, 1428 Buenos Aires, Argentina}
  \affiliation{CEILAP, CITEDEF, J.B. de La Salle 4397, 1603 Villa Martelli, Buenos Aires, Argentina}

\author{Juan Pablo \surname{Paz}}
  \affiliation{Departamento de F\'{\i}sica, FCEyN, UBA, Pabell\'on 1, Ciudad Universitaria, 1428 Buenos Aires, Argentina}
  \affiliation{IFIBA, UBA-CONICET, Pabell\'on 1, Ciudad Universitaria, 1428 Buenos Aires, Argentina}

\author{Christian Tom\'as \surname{Schmiegelow}}
  \affiliation{Departamento de F\'{\i}sica, FCEyN, UBA, Pabell\'on 1, Ciudad Universitaria, 1428 Buenos Aires, Argentina}

\begin{abstract}
Several schemes have been proposed to extend Quantum Key Distribution protocols aiming at improving their security or at providing 
new physical substrates for qubit implementation. We present a toolbox to jointly create, manipulate and measure qubits stored 
in polarization and transverse-modes 
%orbital angular momentum 
degrees of freedom of single photons. The toolbox includes local operations on single qubits, controlled operations between the 
two qubits and projective measurements over a wide variety of non-local
bases in the four dimensional space of states. We describe how to implement the toolbox to perform an extended version of the 
BB84 protocol for this Hilbert space (ideally transmitting two key bits per photon). We present the  experimental implementation 
of the measurement scheme both in the regimes of intense light beams and with single photons.  Thus, we show the feasibility of 
implementing the protocol providing an interesting example of a new method for quantum information processing using the 
polarization and transverse modes
% the orbital angular momentum 
of light as qubits.
\end{abstract}

\maketitle

% Aca describimos los qubits y bases
\section{Introduction}
Quantum Key Distribution (QKD) protocols exploit the quantum non-cloning theorem \cite{no-cloning} and the indistinguishability 
of quantum states belonging to unbiased bases \cite{planat2006survey} to accomplish secure distribution of cryptographic keys. 
After the original BB84 \cite{BB84} protocol, several other protocols have been proposed together with different encoding schemes 
and possible realizations for physical qubits. Different schemes focus on improving the security, the efficiency or on enabling 
practical realizations. In this paper we present a scheme to transmit two key bits on a single photon encoding two qubits on two 
different photonic degrees of freedom. For this we use the polarization and the transverse-modes (TM) degrees of freedom. 
Equivalent schemes were developed earlier with polarization and time-bin qubits \cite{Buttler}. Having access to the complete 
4D Hilbert space enables improvement on the security \cite{security_d_system} and/or the key generation rate of the protocol. 

To implement any QKD scheme it is necessary to prepare and measure any quantum state of a set of mutually unbiased bases (MUBs). 
Thus, such states are the primary resource required to encode and transmit a bit of key.  In this paper we
will show how to prepare and measure any state of a set of mutually unbiased basis in a four dimensional Hilbert space 
(in such space the maximal number of MUBs is five). Any state will be prepared by preparing first a state in a canonical basis 
(the so called computational basis) and then applying a unitary operator to change the basis to the desired one. As mentioned above, 
we will show how to do this with single photons encoding two qubits on polarization and TM degrees of freedom. The unitary 
operators will be implemented by combining simple elementary quantum gates which will be chosen as general operations on both 
individual qubits  and a controlled operation (as it is well known, with these tools we would have universal control on the 
evolution \cite{NielsenChuang}).

This paper is organized as follows. In section \ref{LocalAndControlled} we describe the main ingredients required to understand 
the physics of both polarization and TM qubits. Using them we show how to prepare states of the canonical basis. Finally, in this 
section we show the experimental arrangement that would allow for the implementation of the most general rotation on the Bloch 
sphere of each individual qubit, and a controlled operation between the polarization and the TM degrees of freedom. In 
section \ref{projection} we show how to implement the projective measurement on the canonical basis, which relies on the 
discrimination of TEM$_{01}$ and TEM$_{10}$ modes of light. We show how to implement this with a Mach-Zehnder interferometer 
with an Extra Mirror (MZEM).

Section \ref{4DQKD} is devoted to a description of a general QKD scheme with two key bits transmitted on each individual photon, 
which is an extension of the BB84 protocol to a larger Hilbert space. Also, we discuss potential improvements on the security 
and the key generation rate that could be achieved by this protocol, and we describe the implementation of the protocol for a specific 
maximal set of mutually unbiased basis. Finally, in section \ref{TEM-BStest} we show the results of an experimental implementation of the 
MZEM interferometer used to evaluate in practice the projection onto the canonical basis providing a simple proof of principle of the 
feasibility of the proposed method.

% From this elements a complete scheme for a 4D QKD protocol can be implemented, as will be
% described on section \ref{4DQKD}. The MZEM will be experimentally studied to explore the feasibility of practical implementation
% of the protocol, analyzing the complexity and stability of this part of the setup.

\section{Single-qubit and controlled operations}
\label{LocalAndControlled}

It is well known that for a single qubit any rotation over the Bloch sphere can be obtained from a combination of three rotations
of $\pi$ or $\sfrac{\pi}{2}$ over axes that lie on the same plane. We call them respectively \mbox{$\pi$-converter} and 
\mbox{$\sfrac{\pi}{2}$-converter} \cite{UnivRotator}. On the present implementation for light qubits the axes of rotation
on the Bloch sphere are associated with the angular orientation of the converters on the plane transverse to the propagation axis.

For the polarization qubit we define the states in the computational basis by associating them with horizontal and vertical 
polarization states, i.e. $\ket{0}\equiv\ket{\PolH}$, $\ket{1}\equiv\ket{\PolV}$. These two states are defined as the 
eigenstates of the $Z$ Pauli operator. The eigenstates of the operator $X$ correspond to the diagonal polarization states and 
those of the operator $Y$ are associated with the circular ones. For this choice the \mbox{$\pi$-converter} and 
\mbox{$\sfrac{\pi}{2}$-converter} can be implemented with a Half Wave Plate (HWP) and a Quarter Wave Plate (QWP) respectively. 
Therefore, the most general unitary operator on the polarization qubit can be obtained using the combination 
QWP$(\alpha)\cdot$HWP$(\beta)\cdot$QWP$(\gamma)$, where $\alpha,\beta,\gamma$ are physical rotations of the wave plates around the propagation axis. 
We call this sequence the Universal Polarization Rotator (UPR) \cite{UnivRotatorPipo}.

To build the TM qubit we should take into account that transverse modes are solutions of the paraxial Helmholtz equation for the 
distribution of phase and  amplitude of light over the plane transverse to the propagation axis. There are several families of 
modes that include the Hermite Gaussian modes for Cartesian coordinates and the Laguerre-Gaussian modes for cylindrical coordinates.
These families have been studied both classically \cite{PhysRevA.45.8185,Beijersbergen} and from the perspective of 
quantum optics\cite{vanEnk1992147,vanEck1ro}.
Each family forms a 
complete orthogonal basis of the space of square-integrable functions over the plane, so any mode of a given family can be 
expanded as a linear combination of the modes of any other family. The relation between modes has been obtained, for example, 
in \cite{AritmeticaTEMOAM,PhysRevA.45.8185,Beijersbergen}. To build the Hilbert space of TM states we use each basis of TM modes 
as a possible basis of such state. 

Of course, there are infinite choices of pairs of orthogonal states that can be used to define the TM qubit. In fact, many such 
qubits have been recently studied in the literature \cite{SeparadorModos1,Padgett99}. Here, we define the computational states 
of the TM qubit by identifying $\ket{0}\equiv$TEM$_{01}$ and $\ket{1}\equiv$TEM$_{10}$, where TEM$_{ab}$ are Hermite Gaussian 
modes of order $a+b=1$. These states are defined as eigenstates of the $Z$ Pauli operator of the TM qubit. The eigenstates of 
the $X$ operator are diagonal Hermite Gaussian modes (which are rotated by an angle $\sfrac{\pi}{4}$). In turn, the eigenstates 
of the $Y$ operators are Laguerre Gaussian modes, which are known to carry orbital angular momentum \cite{AritmeticaTEMOAM}. 
In this way we build a Poincar\'e sphere for spatial modes which is expressed 
in Table \ref{tab:basis} and is analogous to the one defined by Padgett \etal \cite{Padgett99}.

%\begin{savenotes}
\begin{table}[h]
 \centering
  \begin{tabular}{|c|c|c|c|}
  \hline
  \mc{4}{|c|}{Bases for TM qubit} \\
  \hline
  Vector \;   & \;$Z$\; & \;$X$\; & \;$Y$ \\
  \hline
  $\ket{0}$\; & \;$\ket{\TEMH}$ \; & \;$\ket{\TEMD}=\frac{\ket{\TEMH}+\ket{\TEMV}}{\sqrt{2}}$\; 
                                   & \;$\ket{\TEMR}=\frac{\ket{\TEMH}+i \ket{\TEMV}}{\sqrt{2}}$ \\[1ex]
  $\ket{1}$\; & \;$\ket{\TEMV}$ \; & \;$\ket{\TEMd}=\frac{\ket{\TEMH}-\ket{\TEMV}}{\sqrt{2}}$\; 
                                   & \;$\ket{\TEML}=\frac{\ket{\TEMH}-i \ket{\TEMV}}{\sqrt{2}}$ \\
  \hline
  \end{tabular}
 \caption{ Description of the three mutually unbiased bases for the TM qubit in terms of Hermite-Gaussian and Laguerre-Gaussian 
           modes. The black and white fillings account for the $\pi$ phase shift between lobes of each state. The eigenstates of 
           the $Y$ basis are two Laguerre-Gauss TEM$_{10}$ modes with counter-propagating helical wavefronts}
 \label{tab:basis}
\end{table}

Analogously, in order to define the rotation operators of the TM qubit we can proceed as follows: The \mbox{$\pi$-converter} can be 
physically implemented with two cylindrical lenses separated by a distance that is equal to two focal lengths (properly 
mode-matched to the incoming beam).  Similarly, a \mbox{$\sfrac{\pi}{2}$-converter} can be realized with two cylindrical 
lenses separated by a distance equals to $\sqrt{2}$ times the focal length \cite{Beijersbergen}. Therefore, using a careful 
alignment of three cylindrical lens pairs at the appropriate angles, we can implement any rotation of the TM qubit. We denote 
this arrangement as a Universal Transverse-modes Rotator (UTR).

\begin{figure}[htb]
 \centering
U \includegraphics[width=0.45\textwidth,keepaspectratio=true]{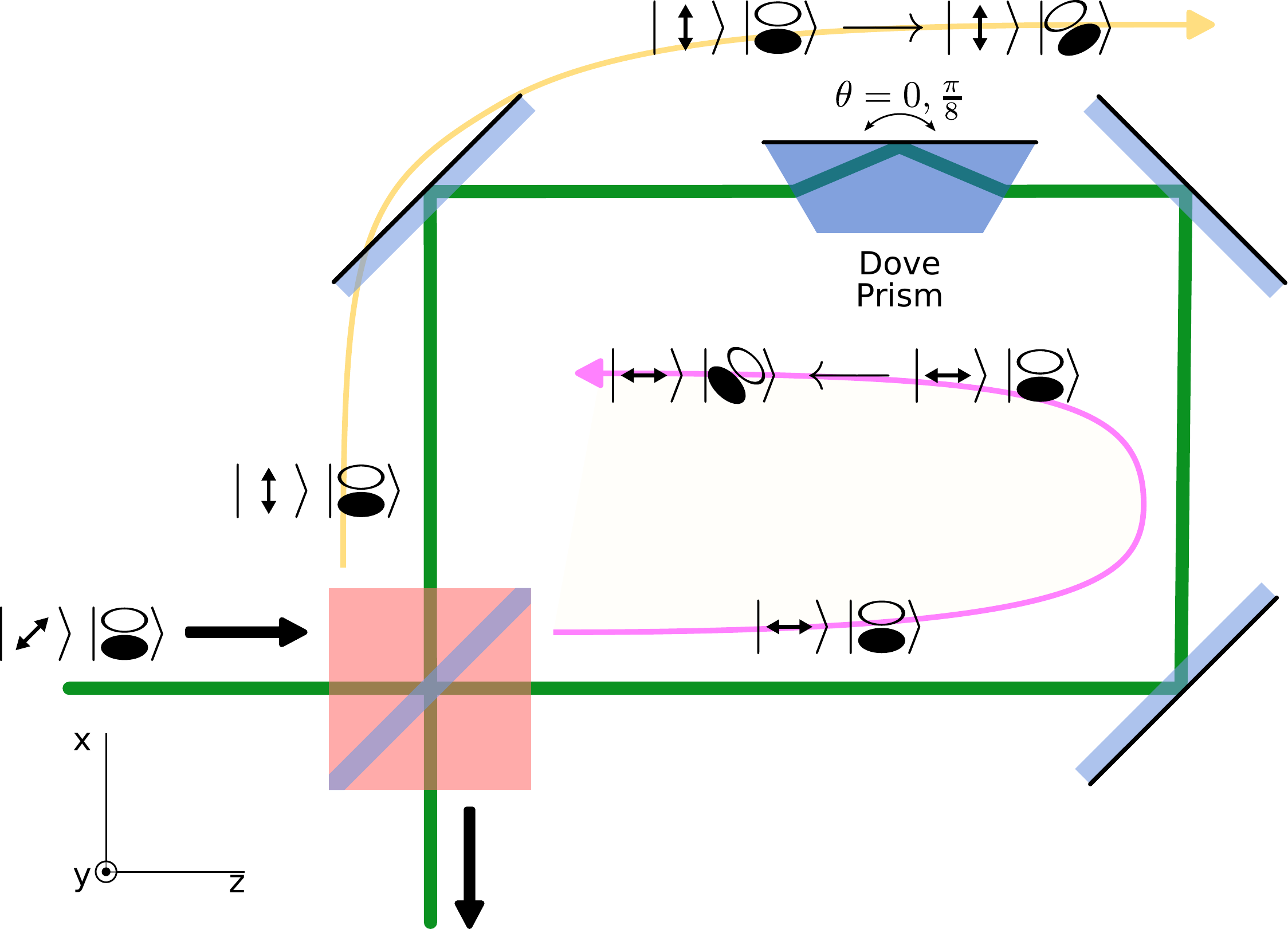}
 % sagnac.pdf: 647x467 pixel, 72dpi, 22.82x16.47 cm, bb=0 0 647 467
 \caption{Sagnac interferometer with a PBS, which implements a controlled operation from polarization to transverse modes. The
           Dove prism applies a rotation to the transverse wavefront distribution, being the sign of this rotation determined 
           by the direction of propagation.}
 \label{fig:sagnac}
\end{figure}

Finally, to complete the set of operations we propose to implement the controlled operation using a Sagnac-type interferometer 
with a Polarization Beam Splitter (PBS)  (Fig. \ref{fig:sagnac}), based on the study of momentum and polarization entanglement 
reported by Fiorentino \etal \cite{CNOT_OAM}. This is a robust implementation because there is no need for dynamical compensation 
of the optical path, being the Sagnac a common-path interferometer. In this setup, by means of a Dove prism which is rotated 
along the propagation axis (a \mbox{$\pi$-converter}), the transverse modes can be rotated by an angle whose sign depends on the 
direction of propagation. In turn, as a consequence of the presence of the PBS, each polarization component travels through the 
interferometer in different directions. Then, the device transforms each polarization in a different way. If the Dove prism is 
positioned as in figure \ref{fig:sagnac}, neither the polarization nor the transverse modes are affected, 
implementing the identity operator. However, if the prism is rotated at an angle of $\sfrac{\pi}{8}$,     
an $U$ operator is applied to the TM qubit for the component that travels in one direction and the $U^\dag$ is applied to the other     
\eqref{eq:sagnac}. Therefore, the change of basis of the TM qubit is controlled by the polarization qubit and can be switched on/off 
by choosing the Dove prism angle. 

\begin{equation}
\label{eq:sagnac}
\begin{aligned}
\text{Sagnac}\left( \frac{\pi}{8} \right) \text{:} & \\ &
  \Qcircuit @C=1em @R=.7em { 
  \lstick{\ket{\PolH / \PolV}} &  \ctrlo{1} & \ctrl{1}     & \qw \\
  \lstick{\ket{\TEMH / \TEMV}} &  \gate{U\phantom{^\dag}} & \gate{U^\dag} & \qw 
  }
\quad
  \text{U}=
  \frac{1}{\sqrt{2}}
  \begin{bmatrix}
  \phantom{-}1 & 1 \\
	  - 1 & 1 
  \end{bmatrix}
\end{aligned}
\end{equation}

Combining the previous operations we can change basis from the computational one to any other basis in the Hilbert space. 
Using the UPR and UTR we build any local transformation. Combining them with the controlled operation, we can prepare any 
state starting from the computational basis. This also enables measurement on any basis since we can first rotate to 
the computational one and then perform a projective measurement on the canonical basis.

\section{Projection on the canonical basis}
\label{projection}
For the polarization qubit the projective measurement onto the computational basis can be easily performed using a PBS. 
This separates the paths of the $\ket{\PolH}$ and $\ket{\PolV}$ components of the incoming state. Then, they can be separately 
detected with single photon detectors. This simple optical element implements the measurement onto the $Z \otimes I$ basis of 
the polarization qubit.

\begin{figure}[htp]
 \centering
 \includegraphics[width=0.45\textwidth,keepaspectratio=true]{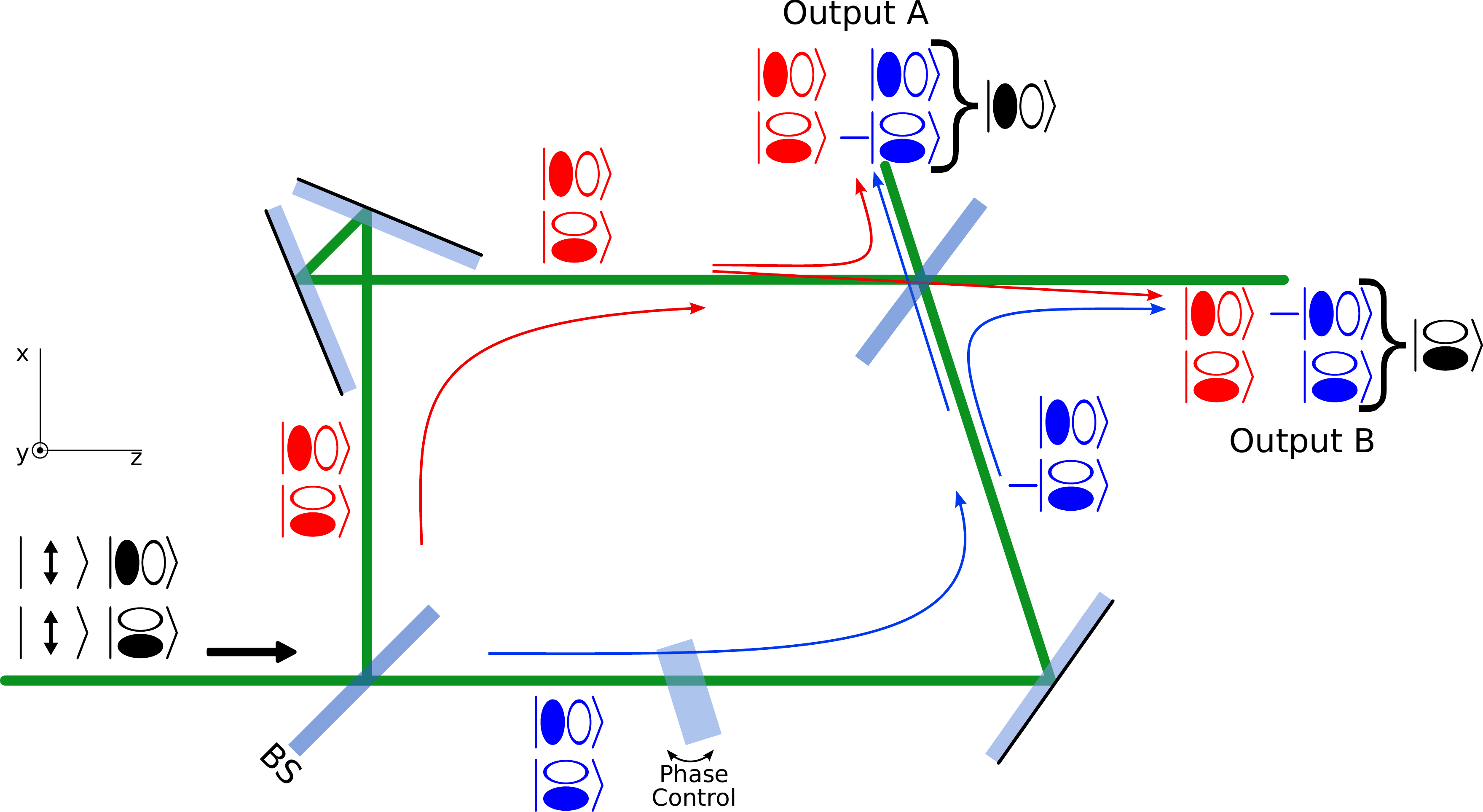}
 % MZespejoextra.pdf: 1013x555 pixel, 72dpi, 35.74x19.58 cm, bb=0 0 1013 555
 \caption{Mach-Zehnder interferometer with an Extra Mirror, that can discriminate $\ket{\protect\TEMH}$ and $\ket{\protect\TEMV}$ 
          states. The states with symmetry against specular reflection along the $\hat{y}$ axis remain invariant after mirror 
          reflections, and the states with anti-symmetry get a $\pi$ phase. The superposition on the MZEE output generates 
          destructive and constructive interference on the complementary exits for sates with different parity eigenvalues.}
 \label{fig:MZEE}
\end{figure}

The measurement onto the eigenstates of the $Z$ basis associated with the TM qubit can be performed with the same idea. We simply 
need a TM beam splitter that splits up the beam into its $\ket{\TEMH}$ and $\ket{\TEMV}$ components. This can be implemented 
using a Mach-Zehnder interferometer with an Extra Mirror as shown in Fig. \ref{fig:MZEE}, as proposed by Sasada \etal 
\cite{SeparadorModos2}. Such interferometer works by exploiting the symmetry (anti-symmetry) of modes $\ket{\TEMH}$ ($\ket{\TEMV}$) 
against specular reflection along the $\hat{y}$ axis. On each output mode of the MZEM there is interference between the incoming 
wavefront and its mirror reflection. Because of the extra mirror there is an additional phase and the system is therefore 
selective to states with vertical symmetry. Thus, interference is constructive on one output while it is destructive in the 
other one, depending on the parity of the input state. A phase plate inserted in one of the arms of the MZEM allows for 
additional control over the interference condition at the outputs, as in a standard amplitude division interferometer.

The two states \{$\ket{\PolH}$,$\ket{\PolV}$\} have different parity. As the MZEM applies the operator $Z \otimes Z$, the output 
depends on the TM state \emph{and} on the polarization state of the incoming photon. The correct identification of the projected 
TM qubit strongly depends on the optical path difference between the arms of the interferometer. Sub-wavelength stability is 
necessary and, as a consequence the MZEM requires dynamic optical path stabilization. Taking this into account, the different 
TM modes can be discriminated with the above device.

Using the projective measurement onto the $Z \otimes Z$ and the $Z \otimes I$ basis, any state of the full 4D computational basis 
can be identified (since each one of them corresponds to one of the four different pairs of measured eigenvalues). We will 
use this measurement scheme in the 4D QKD method, which is described in the following two sections. 
 
\section{Extended QKD with polarization and transverse modes}
\label{4DQKD}

The protocol we present here is an extension of the BB84 protocol \cite{BB84} for a  larger 4D Hilbert space. For this we use two 
or more mutually unbiased bases to encode and measure two bits. In the next subsection we describe this 
4D QKD procedure in abstract terms and in the following section we describe the actual procedure for implementing it using the 
polarization and TM states of a single photon.

\subsection{The general QKD protocol for two qubits}
The protocol is as follows: Alice encodes two random bits by choosing one of the four states of a basis which is chosen at 
random from a set of five mutually unbiased bases. She sends the state to Bob who performs a projective measurement on a basis 
which is randomly chosen from the same set of MUB's. 
After repeating this round of quantum communication many times, they publicly announce the sequence of bases they used. When the 
basis chosen by Alice and Bob coincide they get a shared pair of random bits. With this in hand, they can distill a secure sequence of 
random bits using classical algorithms for error correction \cite{InformationReconciliation94} and privacy amplification 
\cite{PrivacyAmplification}.

If an intruder, Eve, tries to steal information from the key, she would have to perform measurements on a randomly chosen basis 
and resend the quantum state to Bob. This would generate distortions on the correlations between Alice and Bob. Thus, Alice and 
Bob can detect these distortions and ensure the privacy of the key with privacy amplification algorithms. Such task can be done only 
if the quantum bit error rate induced by Eve's attack is not greater than a  critical value of tolerance. This value depends 
on the dimension of the Hilbert space, on the number of bases used on the protocol and on the kind of attacks Eve can perform. 
This has been studied in previous works based on the BB84 protocol \cite{security_2_system} and also for extended protocols 
\cite{security_d_system}. By encoding several qubits on a single photon it could be possible to enhance the level of 
tolerance for the quantum bit error rate, using only two mutually unbiased bases to encode the bits (these two bases can 
be selected from any maximal set of five mutually unbiased bases that can be build for a 4D Hilbert space \cite{MUBS-dimension}). 
If the entire set of five mutually unbiased bases is used, the tolerance to errors can be further increased by a small 
amount \cite{security_d_system}. It is worth mentioning that there is a tradeoff between this improvement in security with 
five MUBs and the increase in the rate of key bit generation that is maximal when one uses only two basis, as was noted in 
\cite{Buttler}.

There is an infinite number of sets of five mutually unbiased bases that can be chosen to implement the above protocol. 
However, there are sets which are particularly simple as they arise as eigenstates of commuting sets of generalized Pauli 
operators. One can always choose a set of three basis which are separable. In fact, they are formed by the eigenstates of 
$X$, $Y$ and $Z$ for each individual qubit. To get the D+1=5 mutually unbiased basis one needs to add two bases formed with 
entangled states. Out of the many possibilities to define these bases, here we will choose the set of the Table \ref{tab:bases}.

\begin{table}[h]
 \centering
  \begin{tabular}{|c|c@{,}c@{,}c|}
  \hline
  Basis & \mc{3}{c|}{CSCO} \\
  \hline
  B1\; & \;$ZZ$\; & \;$ZI$\; & \;$IZ$ \\
  B2\; & \;$XX$\; & \;$XI$\; & \;$IX$ \\
  B3\; & \;$YY$\; & \;$YI$\; & \;$IY$ \\
  \hline
  B4\; & \;$YX$\; & \;$XZ$\; & \;$ZY$ \\
  B5\; & \;$XY$\; & \;$YZ$\; & \;$ZX$ \\
  \hline
  \end{tabular}
 \caption{ 
  Bases names and the Complete Set of 
  Commuting Operators that define them, expressed as tensor products between Pauli matrices and the identity operator $I$.
  B1 is the Canonical basis. B4 and B5 are the entangled bases.
 }
 \label{tab:bases}
\end{table}

\subsection{Physical implementation with two qubits per photon}
\label{QKDwithPolTEM}

The toolbox we presented above enables the manipulation and measurement in the full 4D space of states of the two qubits encoded 
in a single photon. This can be used to implement the 4D QKD protocol we described in the previous section. Thus, we implement 
this scheme by encoding two qubits on polarization and TM of a single photon. 

\begin{figure}[htb]
\centering
\includegraphics[width=0.45\textwidth,keepaspectratio=true]{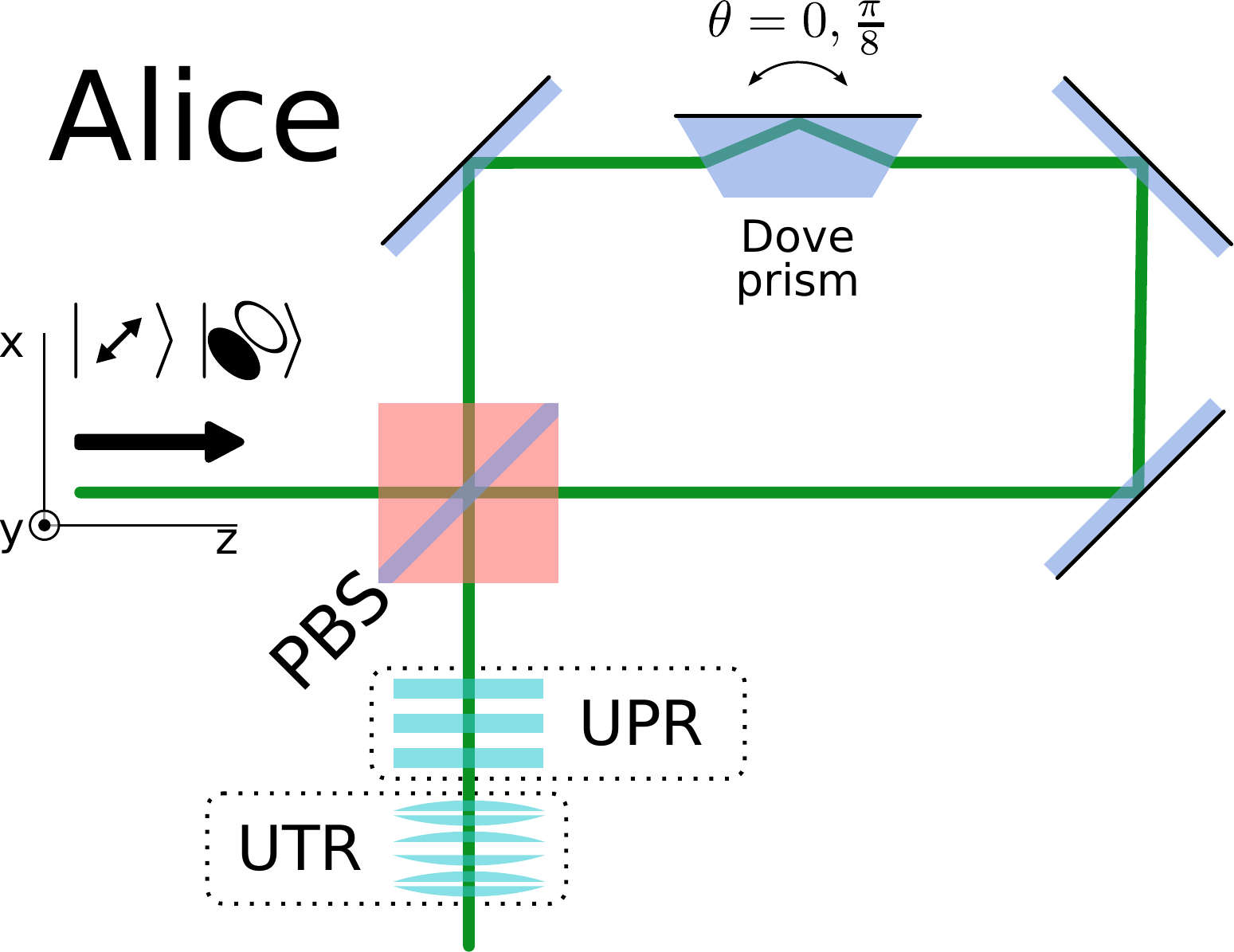}
% MZEEgeo.pdf: 805x471 pixel, 72dpi, 28.40x16.62 cm, bb=0 0 805 471
\caption{Scheme of Alice's setup to prepare a general state using single-qubit operations for each
qubit (UPR and UTR) and a controlled operation implemented with the Sagnac interferometer and a Dove prism.}
\label{fig:alices}
\end{figure}

For this purpose Alice has to prepare an arbitrary state from any of the $D+1=5$  mutually unbiased basis. This is done by using 
an experimental setup as the one shown in Fig. \ref{fig:alices}. The input state is $\ket{\PolD\TEMD}$ from the basis B2. State 
preparation is separated in two parts: The first one consists in the basis choice and 
the second step involves the choice of a state within the chosen basis. Therefore, Alice first chooses whether to prepare a state 
either from the entangled or the product bases sets. For any state within the product basis set she must switch off the the controlled 
operation, which conversely it must be turned on if the chosen state corresponds to the entangled set. Then she decides which basis is 
finally chosen within the selected set by applying single-qubit operations. Also, with single-qubit operations she defines which of the four 
states of the basis is prepared. The detailed description of the state preparation is expressed as sequences of operators 
in the Appendix \ref{append}.

\begin{figure}[htb]
 \centering
 \includegraphics[width=0.45\textwidth,keepaspectratio=true]{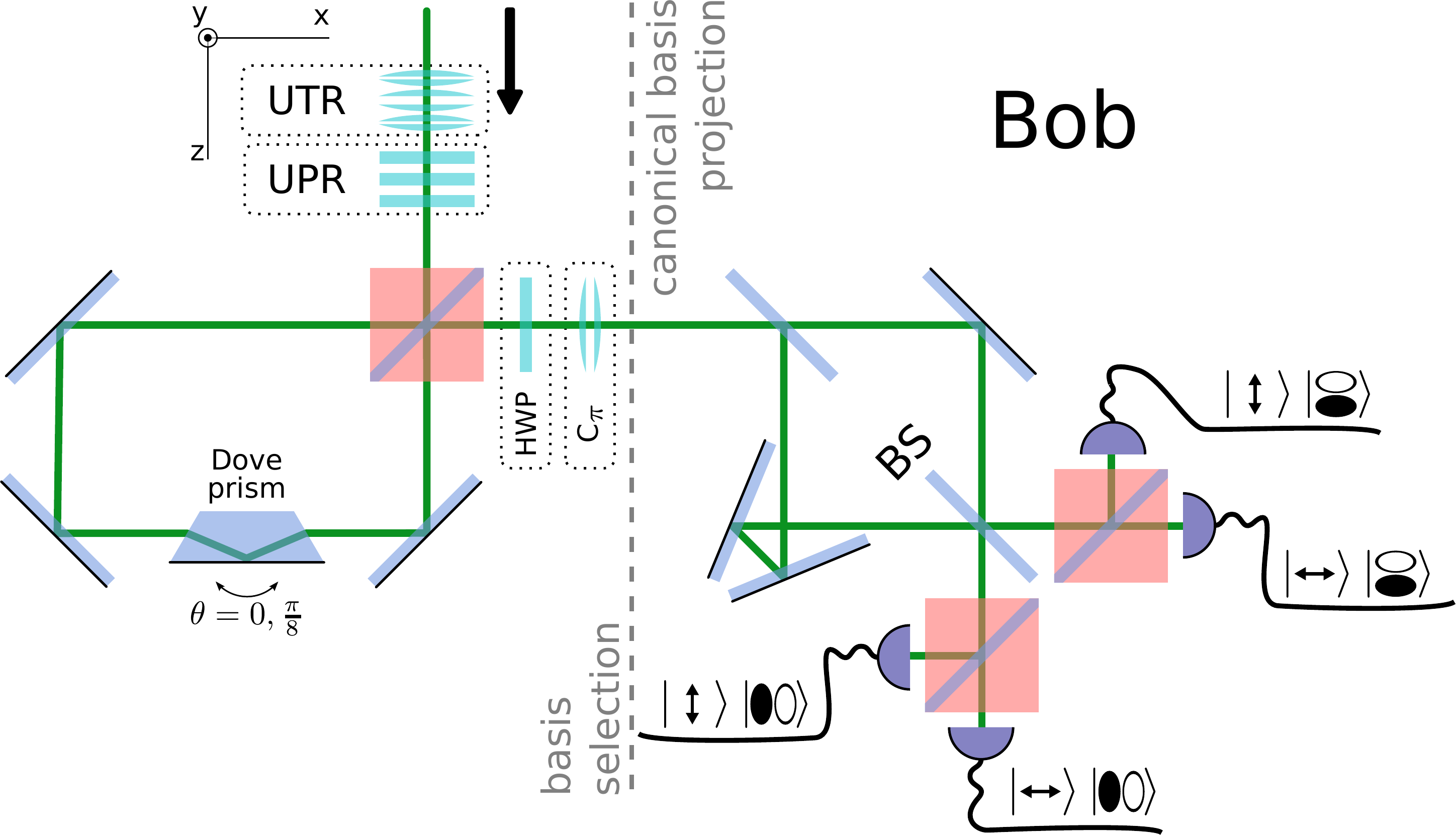}
 % MZEEgeo.pdf: 805x471 pixel, 72dpi, 28.40x16.62 cm, bb=0 0 805 471
 \caption{Scheme of Bob's setup, with which any basis projection can be obtained. The ``basis selection'' part consists on a 
 copy of reversed copy of Alice's setup and an extra $\pi$-converter, which can map any basis onto the canonical one. The 
 ``canonical basis projection'' part consists on the combination of a TM discriminator (MZEM), two polarization
 discriminators (PBS) and four single-photon detectors that implement a projective measurement of the canonical basis.}
 \label{fig:bobs}
 \end{figure}

At the other end, Bob must perform a projective measurement onto any of the possible bases. For this he has to apply the 
inverse procedure used by Alice to prepare the corresponding basis. In this way he would map the chosen basis onto the basis 
B2. This basis must be further rotated to B1, the computational basis, using a HWP and a $\pi$-converter. Finally, to detect 
a single state within this basis he must use a MZEM, two PBS's and 4 different photon detectors. In this way he completes 
the projective measurement (Fig. \ref{fig:bobs}).  %\textcolor{red}{CHECK}

For a correct implementation of the protocol, the optical alignment of all the optical elements must be fine-tuned. This is not 
straightforward because the overall efficiency of the method is limited by factors such as: the correct mode-matching at the 
cylindrical lenses, the losses imposed by reflections on optical surfaces and the efficiency of qubit discrimination. 
Off-the-shelf PBS with low losses and high extinction ratios for a defined wavelength can be used, but the efficiency of 
the MZEM depends on several other factors that will be analyzed in the next section.

\section{TM Beamsplitter test}
\label{TEM-BStest}

We built and tested a MZEM to evaluate the feasibility of using TM of light as qubits using the above scheme. We were able 
to generate and discriminate all the canonical basis states using the proposed device. The results we present here are similar 
to those recently reported by Sasada \etal \cite{SeparadorModos2} for the case of intense light beams. However, here we show 
the performance of the method to discriminate TM on the single-photon limit, which is essential for a QKD protocol. In what 
follows we describe the setup and the techniques we used to obtain good visibility on the interferometer, which is a 
mandatory condition to realize QKD.

The efficiency of the MZEM strongly depends on the quality of the TM. As described above, a different interference pattern 
is produced depending on the properties of the modes under reflection in the mirror. Then, the device is very sensitive to defects 
on the reflection symmetry of the transverse intensity profile. It is also very sensitive to the preservation of the symmetry or 
anti-symmetry of the phase distribution. Thus, the initial TM state preparation has to be done carefully in order to control 
these features.

\begin{figure}[ht]
 \centering
  \includegraphics[width=0.45\textwidth,keepaspectratio=true]{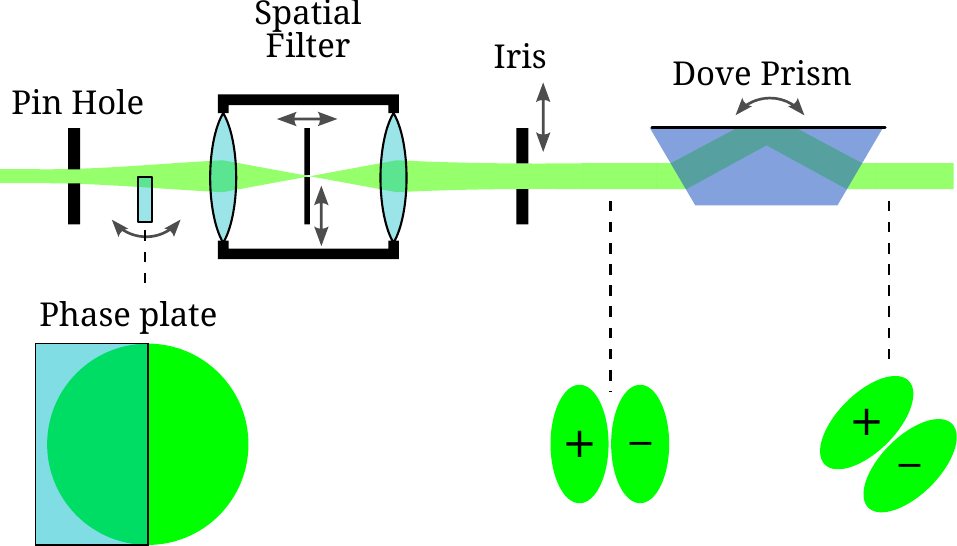}
 \caption{Scheme of the TEM builder used to test the MZEM. The lower part shows a simplified picture of the beam profile obtained 
          at different stages of the process}
 \label{fig:modes_generation}
\end{figure}

The photon source was built from a CW diode-pumped, intracavity doubled Nd-YAG laser, using a BBO crystal which was cut for 
type-II phase-matching condition. This system emits a few milliwatts of a 532 nm beam with an effective coherence length of 5 mm.
The TM modes were built generating a $\pi$ phase change over half of the wavefront using a thin glass plate. The resulting beam 
was spatially filtered and collimated. An iris diaphragm was used to control the spot size and a Dove prism 
allowed for rotation of the transverse pattern around the propagation direction (Fig. \ref{fig:modes_generation}), in order 
to select the transverse-mode state of the canonical basis. Finally, the polarization state was selected with a PBS. 
When needed, the beam was attenuated with neutral density filters and a Malus-like device to achieve the single photon regime. 

\begin{figure}[ht]
 \centering
  \includegraphics[width=0.3\textwidth,keepaspectratio=true]{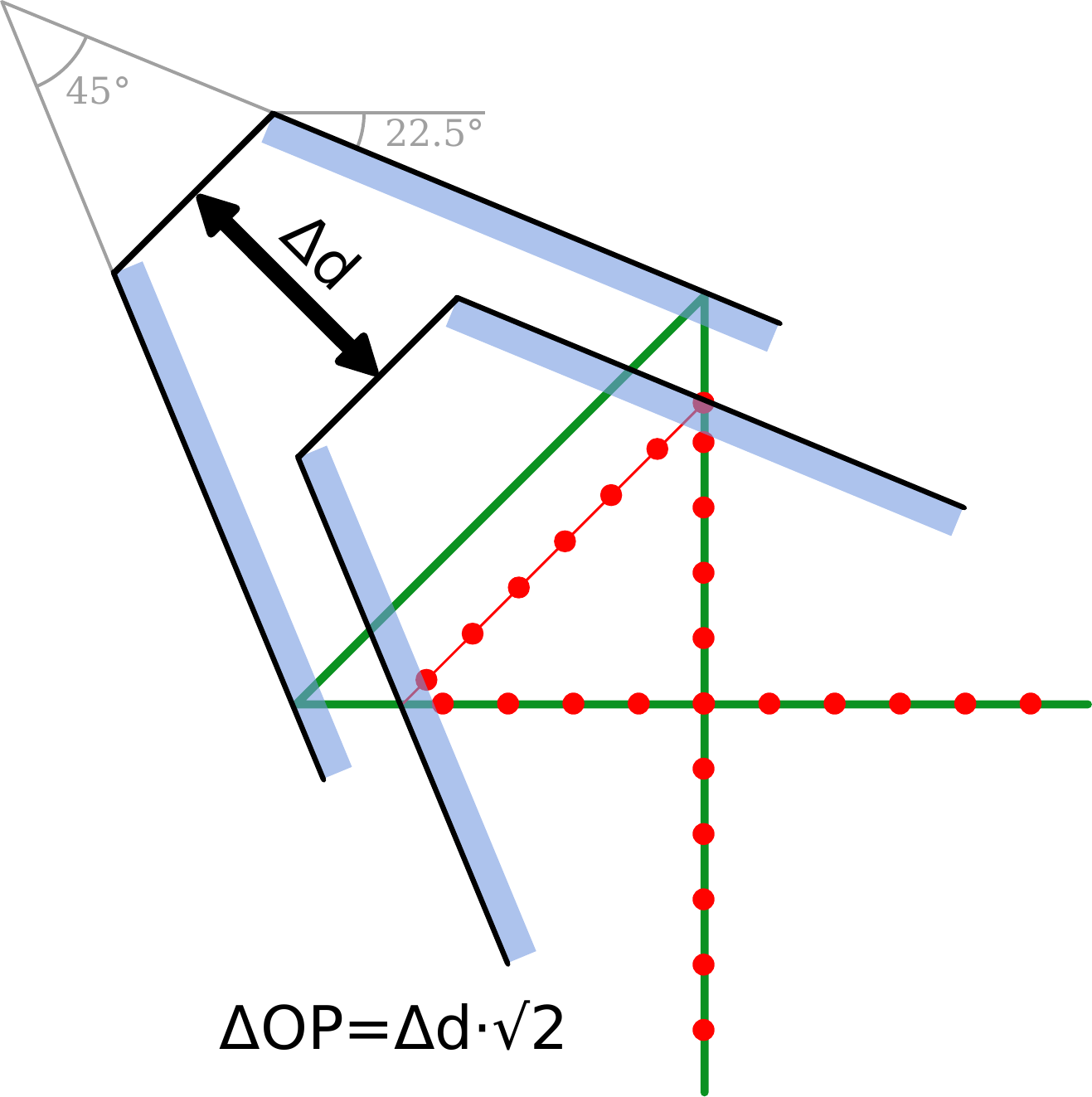}
 \caption{The geometrical properties of the double mirror enables to generate a change over the optical path length. 
          A displacement of $\Delta d$ over the symmetry axis produces a difference of $\sqrt{2} \,\Delta d$ on the 
          optical path.}
 \label{fig:pentaprisma}
\end{figure}

Construction of the MZEM requires a more careful alignment than a standard Mach-Zehnder interferometer. One key issue is the 
double mirror reflection, that introduces an additional optical path on one of the arms that has to be compensated on the 
other one. The position of the single mirror (Fig \ref{fig:MZEE}) was estimated by simple geometric calculations. The fine 
tuning of the path compensation was realized  by mounting the double mirror assembly on a micrometer-driven translation 
stage and scanning it along its symmetry axis, exploiting geometrical properties of the design (Fig. \ref{fig:pentaprisma}). 
An auxiliary laser source with $0.1$ mm coherence length was used to find the optimal position of the double mirror position 
to reduce the path difference below $0.1$ mm, achieving optimal visibility for TEM$_{00}$ modes.

The optical path was actively compensated during measurements using a He-Ne laser aligned collinear with the photon source 
path and vertically displaced from it. The temporal behavior of the He-Ne interference pattern was registered with a photodiode 
and used as source of a feedback circuit to correct the position of a single mirror mounted over a piezoelectric actuator, 
thus locking the optical path difference with an estimated accuracy of 10 nm.

The MZEM was tested for the four states of the canonical basis, both on an intense beam regime and on a weak beam regime, i.e. 
the single photon regime. Table \ref{tbl:resultados_HV} shows the high intensity transverse modes at the output of the MZEM 
for the canonical basis states, registered with a CCD camera and colored by intensity. A particularly sound demonstration of the MZEM 
operation is the decomposition of a diagonal state in its canonical components 
(fourth row of Table \ref{tbl:resultados_HV}). Also, the PBS-like behavior described in \cite{SeparadorModos2} was confirmed.

\begin{table*}[hbt]
{
 \centering
\newcolumntype{V}{>{\centering\arraybackslash} m{.13\textwidth} }
\begin{tabular}{|lVVVV|}
\hline
\multirow{2}{*}{ \protect\includegraphics[width=0.09\textwidth,keepaspectratio=true]{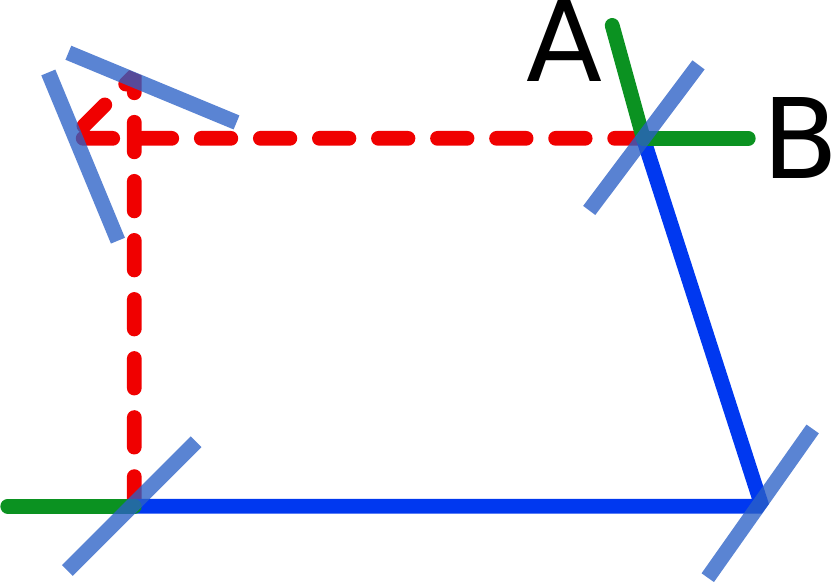}} 
 & \mc{2}{c}{Path} & \mc{2}{c|}{Interference} \\
 & {\color{red}red (dotted)} & {\color{blue}blue (straight)} & $\phi=0$ & $\phi=\pi$ \\
\shortstack{$\Ket{\PolV}\Ket{\TEMV}$\\Output A}&
\includegraphics[width=0.13\textwidth,keepaspectratio=true]{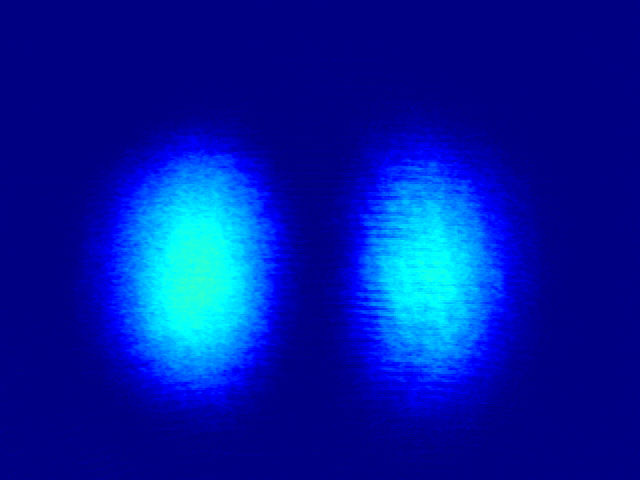} & 
\includegraphics[width=0.13\textwidth,keepaspectratio=true]{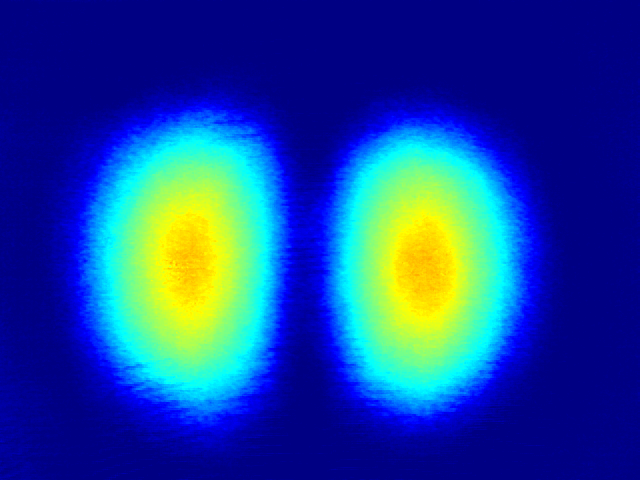} & 
\includegraphics[width=0.13\textwidth,keepaspectratio=true]{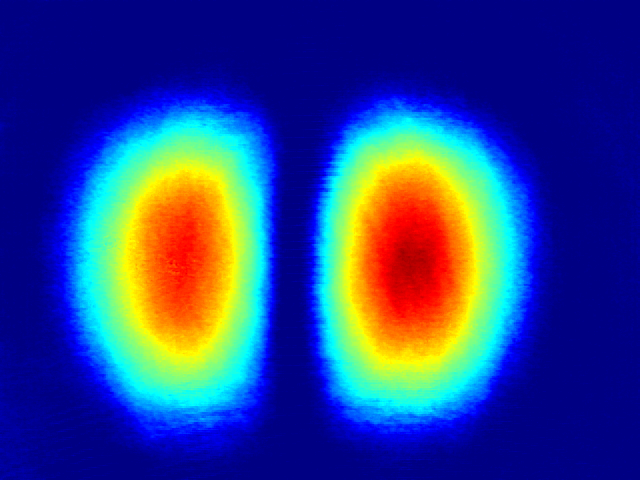} & 
\includegraphics[width=0.13\textwidth,keepaspectratio=true]{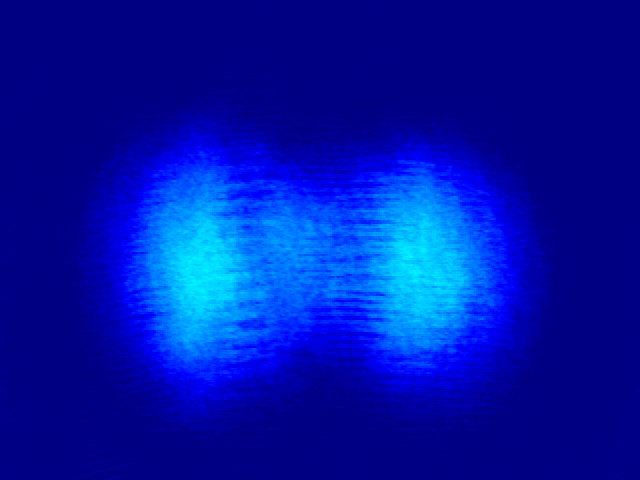} \\
\shortstack{$\Ket{\PolV}\Ket{\TEMV}$\\Output B}&
\includegraphics[width=0.13\textwidth,keepaspectratio=true]{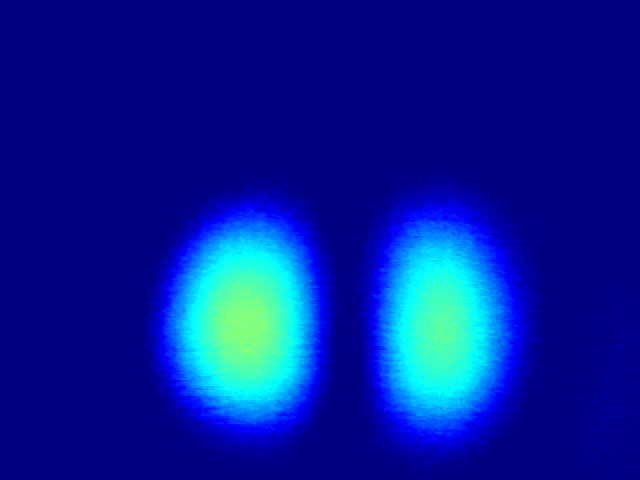} & 
\includegraphics[width=0.13\textwidth,keepaspectratio=true]{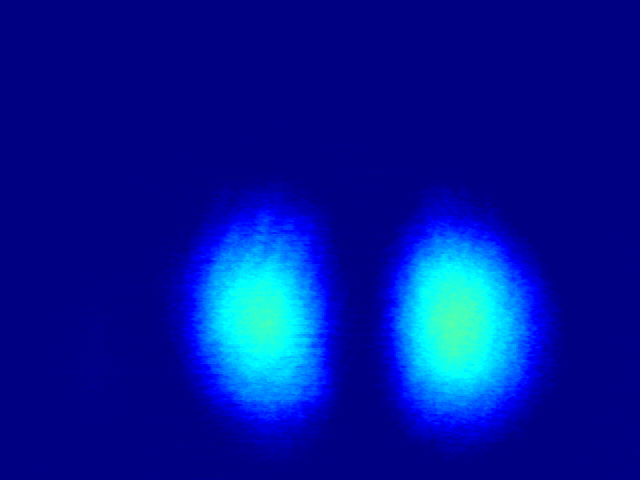} & 
\includegraphics[width=0.13\textwidth,keepaspectratio=true]{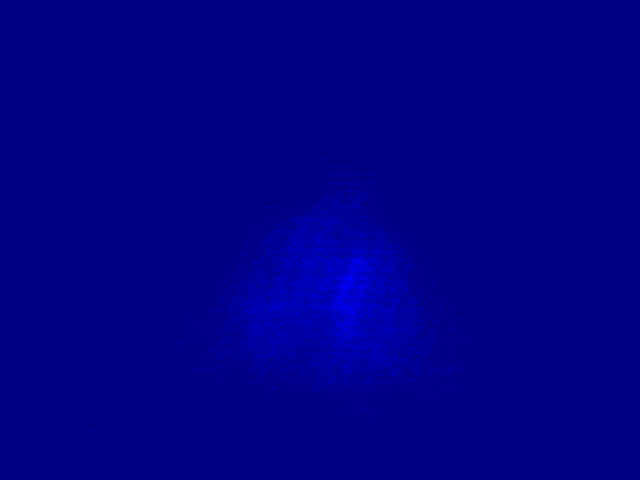} & 
\includegraphics[width=0.13\textwidth,keepaspectratio=true]{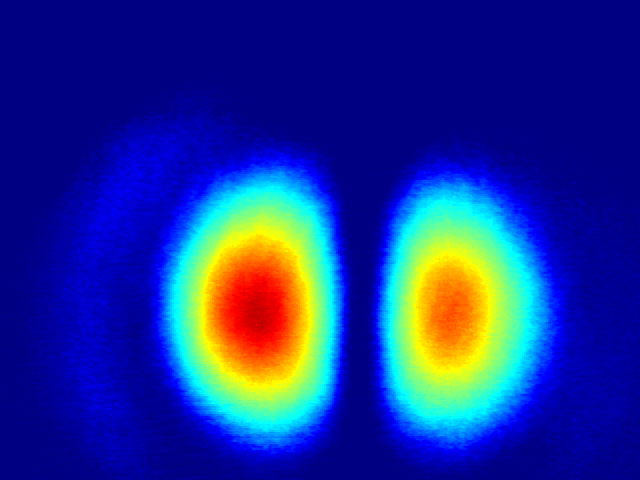} \\
\shortstack{$\Ket{\PolV}\Ket{\TEMH}$\\Output A}&
\includegraphics[width=0.13\textwidth,keepaspectratio=true]{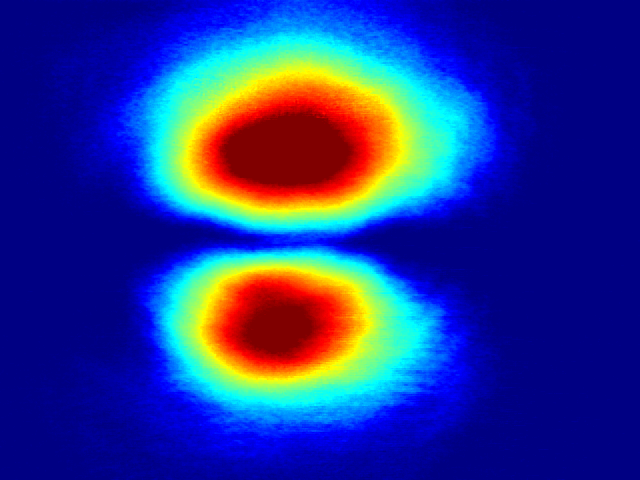} & 
\includegraphics[width=0.13\textwidth,keepaspectratio=true]{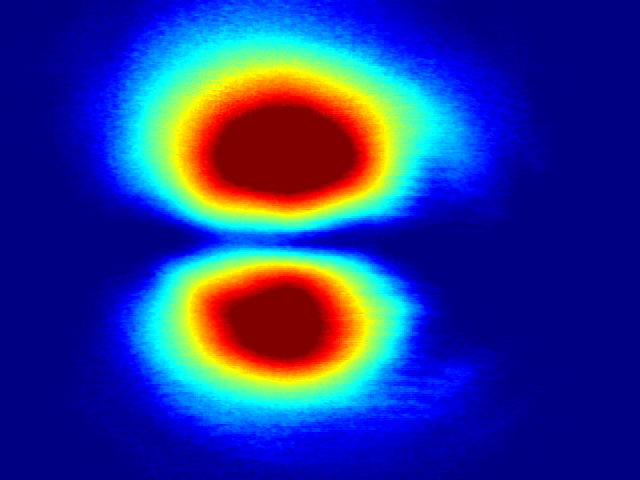} & 
\includegraphics[width=0.13\textwidth,keepaspectratio=true]{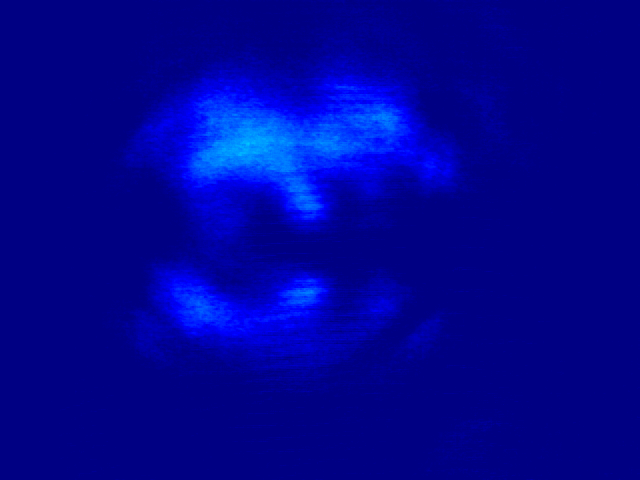} & 
\includegraphics[width=0.13\textwidth,keepaspectratio=true]{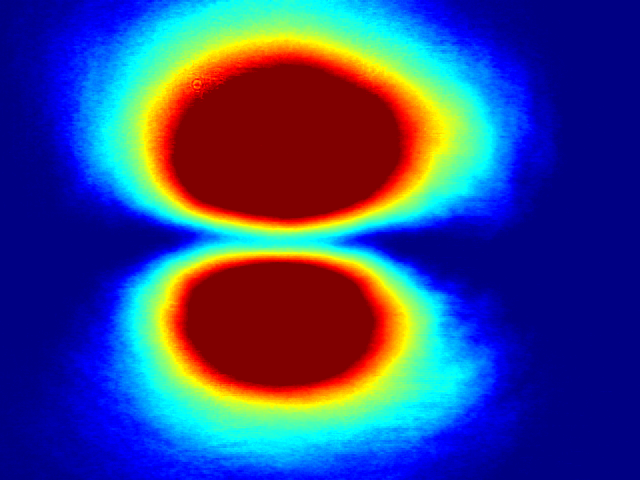} \\
\shortstack{$\Ket{\PolV}\Ket{\TEMD}$\\Output B}&
\includegraphics[width=0.13\textwidth,keepaspectratio=true]{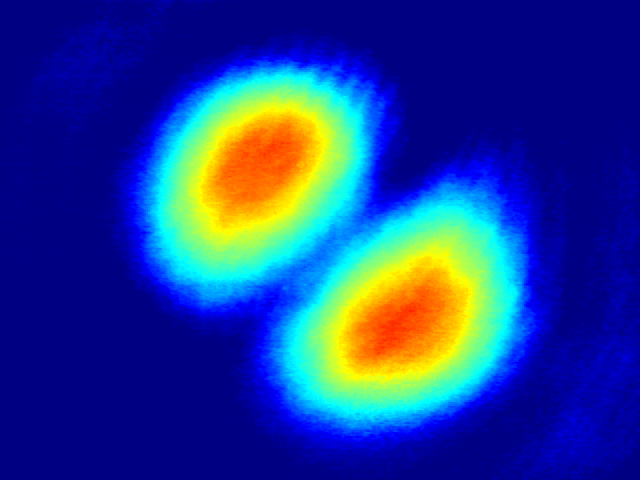} & 
\includegraphics[width=0.13\textwidth,keepaspectratio=true]{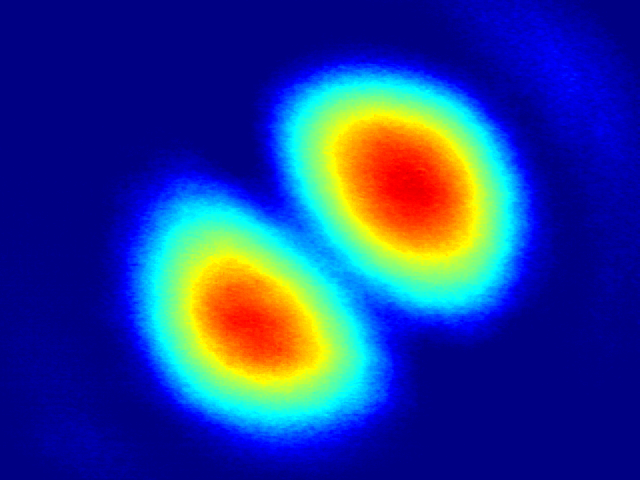} & 
\includegraphics[width=0.13\textwidth,keepaspectratio=true]{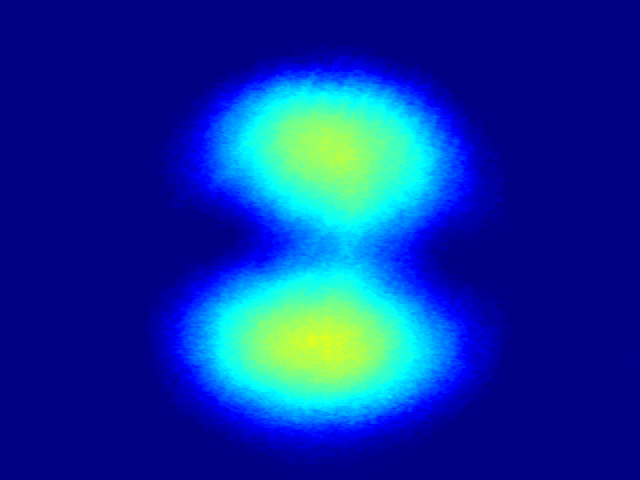} & 
\includegraphics[width=0.13\textwidth,keepaspectratio=true]{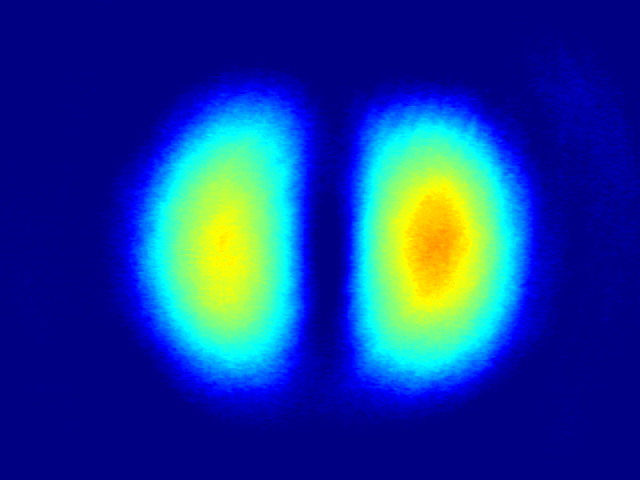} \\
\hline
\end{tabular}
}
\caption{Recorded intensity profiles of different states from the Canonical basis, obtained at the A and B MZEM outputs. 
         On each row, the input is the 2-qubit state at the left. The first two columns correspond to the output beam profiles when one or the other 
         interferometer arms are blocked. The last two columns show the interference patterns for two different (complementary) phase conditions.
         $\phi$ is the relative phase between 
        the two interferometer arms.}
\label{tbl:resultados_HV}
\end{table*}

The performance was also tested at the single photon regime, using a strongly attenuated beam and detecting individual
photons with a set of PhotoMultiplier Tubes (PMTs) Hamamatsu H5783P. The intensity was adjusted to achieve a count rate that 
guarantees a photon rate lower than one photon per transit time at the MZEM, as a condition for a single photon regime.
PMTs were selected over Avalanche Photodiodes because they have a larger detection area that enables the collection of the 
entire spot of the spatial modes of light.

For the single photon regime, the interference visibility over both outputs of the MZEM was 
measured for each state of the canonical basis using photon counting. The results (Table \ref{tab:single-photons-results})
show a visibility of $V\approx0.9$, which is sufficient to perform a proof-of-concept demonstration of a 4D QKD protocol.

\begin{table}[ht!]
 \centering
 \begin{tabular}{|l|c|c|c|c|}
  \hline
         & $\ket{\PolH,\TEMH}$ & $\ket{\PolH,\TEMV}$ & $\ket{\PolV,\TEMH}$ & $\ket{\PolV,\TEMV}$\\[0.5ex]
  \hline
  Output A & $95\pm8$ & $91\pm 13$ & $65\pm 12$ & $68\pm 9$\\[0.5ex]
  Output B & $98\pm7$ & $95\pm 12$ & $83\pm 18$ & $75\pm 16$\\
  \hline
  \end{tabular}
 \caption{Visibility (\%) of each exit measured for every input state of the canonical basis on the single photon regime.}
 \label{tab:single-photons-results}
\end{table}

\begin{figure}[h]
 \centering
 \includegraphics[width=0.45\textwidth,keepaspectratio=true]{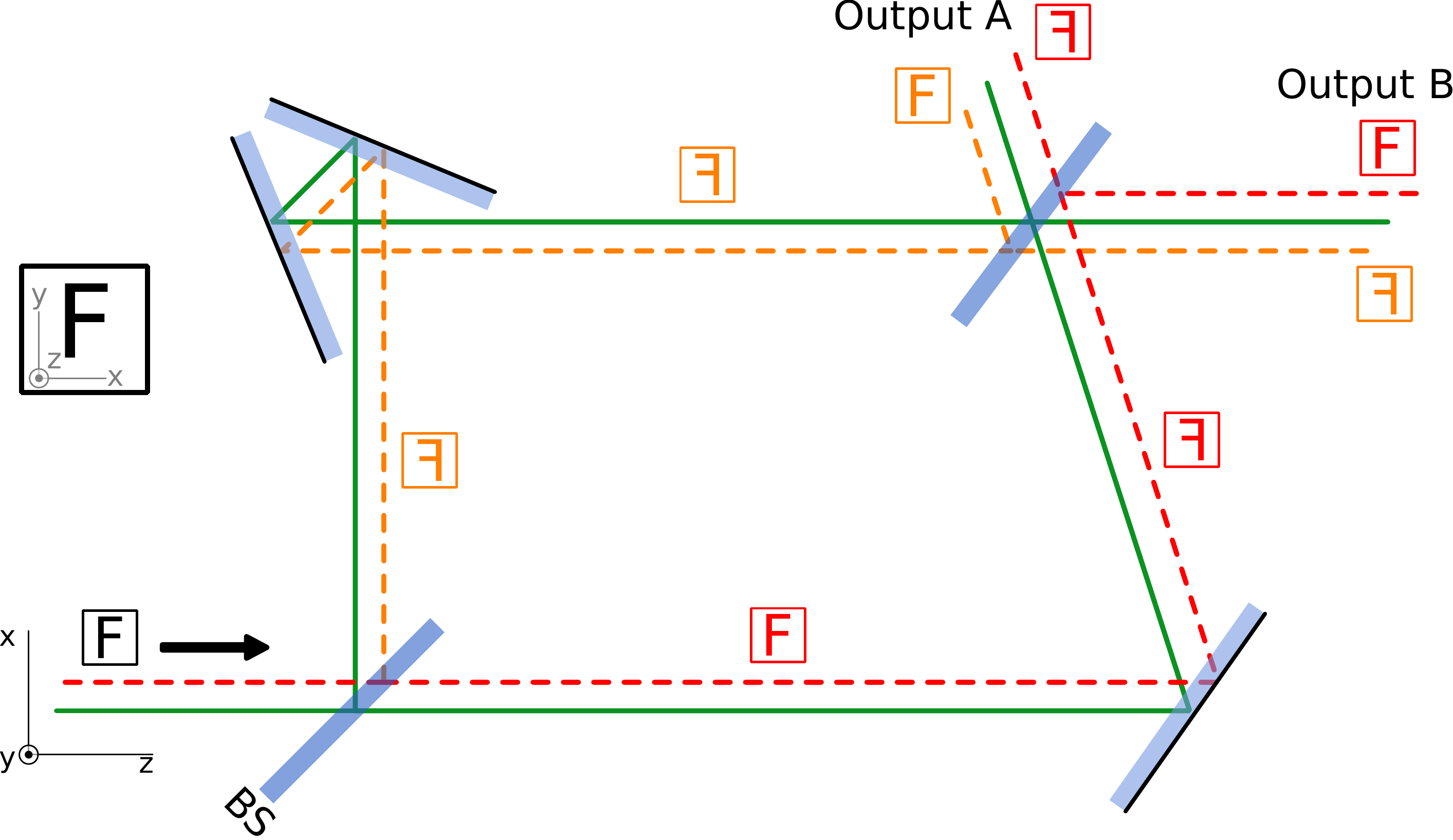}
 % MZEEgeo.pdf: 805x471 pixel, 72dpi, 28.40x16.62 cm, bb=0 0 805 471
 \caption{Geometry considerations for the beam propagation on the MZEM. In contrast to a normal Mach-Zehnder, a lateral 
          shift on the incident beam produces a relative displacement of the two beams at the input of the second beamsplitter, 
          which limits the interference effect to the overlap area between the two displaced beams. On the scheme, an aligned (green) 
          and a displaced (doted red/orange) beam are shown. The F symbol depicts the transformation of the transverse intensity 
          pattern of each beam by the mirror reflections.}
\label{fig:geom}
\end{figure}

The visibility at the output A (Fig. \ref{fig:geom}) is slightly lower because of imperfect BS splitting ratio at the working 
wavelength. The visibility of vertical TM qubits is considerably lower because of an inherent characteristic of the MZEM
alignment:  the interferometer is extremely sensitive to lateral displacements of the incident beam, as  depicted in 
figure \ref{fig:geom}, thus producing a separation over the output beams and deteriorating the interference pattern. 
This feature is particularly detrimental for states that lack vertical symmetry on the intensity profile. These difficulties can 
nevertheless be controlled by careful alignment to achieve good and stable visibility with no fundamental limitations.

 \section{Conclusions}
We presented a toolbox for the implementation of a QKD protocol in a four dimensional Hilbert space. In particular, we studied 
in detail the potential application of this protocol using the  two qubits that can be stored in a single photon using the 
polarization and TM degrees of freedom. The scheme we presented enables us to perform each step of the QKD protocol, with 
complexity that depends on the operation that is to be performed. As any involved free-space optical setup, the complete alignment 
of both Alice and Bob stages 
is not trivial. The use of cylindrical lenses in the UTR stage requires a careful mode matching of the incoming beam 
(and a careful alignment of cylindrical systems is a complex and  troublesome task). The operation of the UPRs can be made 
very fast by  using Polarization Controllers, or combinations of several Pockels Cells. Refractive lenses limit the speed of 
implementation of a UTR, although spatial light modulators used as Fresnel lenses appear as an alternative that can enable fast 
switching.
The MZEM accuracy depends strongly on active stabilization of the optical paths. However, this issue can be easily hurdled 
with standard stabilization techniques. The MZEM is also specially sensitive to $\hat{x}$ displacements from the optimal transverse
position. In contrast, the Sagnac interferometer is very robust and practically insensitive to mid- and long term fluctuatios 
produced by thermal and mechanical stress.

The qubit implementation on the transverse electromagnetic modes of a light beam increases the capacity of single photons to carry 
quantum information, and appears as an alternative to other degrees of freedom such as the linear momentum, or path. The presented 
scheme can be used as a proof-of-concept demonstration of a 4D QKD protocol with 2 qubits encoded on each photon. The use of TM
%modes 
in combination with polarization modes can be also used for alignment-free quantum communication protocols 
\cite{aolita2007quantum}; furthermore the ability to prepare and project any state of any of the 5 Mutually Unbiased Bases for 
the photon polarization-transverse modes qubits may clear the way to perform other original experiments on quantum information.

\section{Acknowledgements}
This work was supported through ANPCyT, CONICET and UBACYT grants. C.T.S. was funded by a CONICET scolarship. 
M.A.L. and J.P.P. are fellows of CONICET.

% \bibliography{/home/augusto/Drafts/Bibitex/refs}

%\bibliographystyle{apsrev4-1.bst}
%\bibliographystyle{apsrev4-1}
\bibliographystyle{unsrtnat}
\bibliography{referencias}

\appendix
\section{State preparation}
\label{append}
The first column of Table  \ref{tbl:TEM_bases_generacion} shows the operations Alice must implement for every basis selection, 
starting from the state $\ket{\PolD}\ket{\TEMD}$.
The second column shows the operations that Alice must perform to select which state from that basis is prepared.
Bob can use the reversed circuit of column one to implement the conversion of any basis to the B2 basis. 
All the local gates can be combined as a unique rotation on the Bloch sphere; this can be implemented with the 
UPR and the UTR on the respective qubits. The action of the Sagnac Controlled Operation consists in setting the 
angle of the Dove prism on $\sfrac{\pi}{8}$, and it can be disabled by setting the angle to zero.

\begin{table*}[H]
\begin{center}
  \renewcommand{\arraystretch}{1.1}

\begin{tabular}{|m{5cm}|cc|}
\hline
\begin{center} Basis selection operation circuit \end{center}
& State selection & Prepared state \\
\hline
%\mc{3}{c}{\color{white}B1}\\
\mc{3}{|c|}{\color{black}B1}\\\hline
\multirow{4}{*}{\quad\quad\quad
  \begin{tabular}{c}
  \Qcircuit @C=1em @R=1em {
  \lstick{\ket{\PolD}} & \gate{\text{Had}} & \rstick{\ket{\PolH}} \qw \\
  \lstick{\ket{\TEMD}} & \gate{\text{Had}} & \rstick{\ket{\TEMH}} \qw
  }\\[7ex]
%   \hline
%   $U\;\text{Had}\otimes\text{Had} \Ket{\PolD,\TEMD}$ 
  \end{tabular}
}&
   $I \otimes I$            & $\Ket{\PolH,\TEMH}$   \\
 & $I \otimes \text{MC}_\pi(\pi/4)$  & $\Ket{\PolH,\TEMV}$   \\
 & $H(\pi/4)   \otimes I$            & $\Ket{\PolV,\TEMH}$   \\
 & $H(\pi/4)   \otimes \text{MC}_\pi(\pi/4)$  & $\Ket{\PolV,\TEMV}$   \\[3ex]
\hline
%%%%%%%%%%%%%%%%%%%%%%%%%%%%% B2
%\mc{3}{c}{\color{white}B2}\\
\mc{3}{|c|}{\color{black}B2}\\\hline
\multirow{4}{*}{\quad\quad\quad
  \begin{tabular}{c}
  \Qcircuit @C=1em @R=1em {
  \lstick{\ket{\PolD}} & \gate{I} & \rstick{\ket{\PolD}} \qw \\
  \lstick{\ket{\TEMD}} & \gate{I} & \rstick{\ket{\TEMD}} \qw
  }
  \\[7ex]
%   \hline
%   $U\;\Ket{\PolD,\TEMD}$
  \end{tabular}
}&
   $I \otimes I$        & $\Ket{\PolD,\TEMD}$   \\
 & $I \otimes \text{MC}_\pi(0)$  & $\Ket{\PolD,\TEMd}$   \\
 & $ H(0)      \otimes I$        & $\Ket{\Pold,\TEMD}$   \\
 & $ H(0)      \otimes \text{MC}_\pi(0)$  & $\Ket{\Pold,\TEMd}$   \\[3ex]
\hline
%%%%%%%%%%%%%%%%%%%%%%%%%%%%% B3
%\mc{3}{c}{\color{white}B3}\\
\mc{3}{|c|}{\color{black}B3}\\\hline
\multirow{4}{*}{\quad\quad\quad
  \begin{tabular}{c}
  \Qcircuit @C=1em @R=1em {
  \lstick{\ket{\PolD}} & \gate{\text{S}} & \rstick{\ket{\PolR}} \qw \\
  \lstick{\ket{\TEMD}} & \gate{\text{S}} & \rstick{\ket{\TEMR}} \qw
  }
  \\[7ex]
%   \hline
%   $U\;Q(0)\otimes Q(0)\Ket{\PolD,\TEMD}$
  \end{tabular}
} &
   $I \otimes I$        & $\Ket{\PolR,\TEMR}$   \\
 & $I \otimes \text{MC}_\pi(0)$  & $\Ket{\PolR,\TEML}$   \\
 & $ H(0)      \otimes I$        & $\Ket{\PolL,\TEMR}$   \\
 & $ H(0)      \otimes \text{MC}_\pi(0)$  & $\Ket{\PolL,\TEML}$   \\[3ex]
\hline
%%%%%%%%%%%%%%%%%%%%%%%%%%%%% B4
%\mc{3}{c}{\color{white}B4}\\
\mc{3}{|c|}{\color{black}B4}\\\hline
\multirow{4}{*}{\quad\quad\quad
  \begin{tabular}{c}
   \Qcircuit @C=1em @R=1em {
   \lstick{\ket{\PolD}} & \multigate{1}{\text{Sagnac}}  &   \qw            &  \qw            &  \qw \\
   \lstick{\ket{\TEMD}} & \ghost{\text{Sagnac}}         &\gate{\text{Had}} &  \gate{\text{S}} &  \qw
   }
  \\[7ex]
%   \hline
%   $U\;I\otimes[Q(0) \text{Had}]\;\Ket{\PolD,\TEMD}$
  \end{tabular}
} &
   $I \otimes I$         & $\frac{1}{\sqrt{2}} (\; \Ket{\PolH,\TEMR} + \Ket{\PolV,\TEML} \;)$   \\
 & $I \otimes \text{MC}_\pi(0)$   & $\frac{1}{\sqrt{2}} (\; \Ket{\PolH,\TEML} + \Ket{\PolV,\TEMR} \;)$   \\
 & $ H(0)      \otimes I$         & $\frac{1}{\sqrt{2}} (\; \Ket{\PolH,\TEMR} - \Ket{\PolV,\TEML} \;)$   \\
 & $ H(0)      \otimes \text{MC}_\pi(0)$   & $\frac{1}{\sqrt{2}} (\; \Ket{\PolH,\TEML} - \Ket{\PolV,\TEMR} \;)$   \\[3ex]
\hline
%%%%%%%%%%%%%%%%%%%%%%%%%%%%% B5
%\mc{3}{c}{\color{white}B5}\\
\mc{3}{|c|}{\color{black}B5}\\\hline
\multirow{4}{*}{\quad\quad\quad
  \begin{tabular}{c}
  \Qcircuit @C=1em @R=1em {
  \lstick{\ket{\PolD}} & \multigate{1}{\text{Sagnac}}  &\gate{\text{S}}   & \qw \\
  \lstick{\ket{\TEMD}} & \ghost{\text{Sagnac}}         &\gate{\text{Had}} & \qw
  }
  \\[7ex]
%   \hline
%   $U\;Q(0)\otimes \text{Had} \; \Ket{\PolD,\TEMD}$
  \end{tabular}
} &
   $I \otimes I$         & $\frac{1}{\sqrt{2}} (\; \Ket{\PolR,\TEMH} + \Ket{\PolL,\TEMV} \;)$   \\
 & $I \otimes \text{MC}_\pi(0)$   & $\frac{1}{\sqrt{2}} (\; \Ket{\PolR,\TEMH} - \Ket{\PolL,\TEMV} \;)$   \\
 & $ H(0)      \otimes I$         & $\frac{1}{\sqrt{2}} (\; \Ket{\PolL,\TEMH} + \Ket{\PolR,\TEMV} \;)$   \\
 & $ H(0)      \otimes \text{MC}_\pi(0)$   & $\frac{1}{\sqrt{2}} (\; \Ket{\PolL,\TEMH} - \Ket{\PolR,\TEMV} \;)$   \\[3ex]
\hline
\end{tabular}
\end{center}
         
\caption{ State preparation procedure. The first column shows the algorithm applied by Alice to select a particular basis, starting
          always from the state  $\Ket{\protect\PolD}\Ket{\protect\TEMD}$ of B2. 
          The product bases only need single qubit gates such as the Hadamard (Had), the Phase gate (S) or the Identity (I), which
          can be implemented by the UPR and UTR. The controlled operation build with the Sagnac interferometer is also needed to 
          produce the entangled bases. Bob can reverse this algorithm to convert any basis to B2.
          The second column shows the specific single qubit operations that must be
          performed by Alice after the basis selection to prepare each of the four states. The $H(\alpha)$ and 
          MC$_\pi(\alpha)$ operators accounts for a HWP and a TM $\pi$-converter respectively, rotated by an 
          angle of $\alpha$. 
          The single qubits operators of the basis selection stage and the final state definition stage can be combined in a unique
          rotation of the Bloch sphere implemented by the UPR and UTR.
          }

          \label{tbl:TEM_bases_generacion}
\end{table*}

\end{document}